\documentclass[rnote]{aa} % for the research notes
\usepackage{graphicx}
\usepackage{natbib}
\usepackage{txfonts}
\usepackage{natbib,twoopt}
\usepackage[breaklinks=true]{hyperref} %% to avoid \citeads line fills
\bibpunct{(}{)}{;}{a}{}{,}             %% natbib format for A&A and ApJ
\makeatletter
  \newcommandtwoopt{\citeads}[3][][]{\href{http://adsabs.harvard.edu/abs/#3}%
    {\def\hyper@linkstart##1##2{}%
     \let\hyper@linkend\@empty\citealp[#1][#2]{#3}}}
  \newcommandtwoopt{\citepads}[3][][]{\href{http://adsabs.harvard.edu/abs/#3}%
    {\def\hyper@linkstart##1##2{}%
     \let\hyper@linkend\@empty\citep[#1][#2]{#3}}}
  \newcommandtwoopt{\citetads}[3][][]{\href{http://adsabs.harvard.edu/abs/#3}%
    {\def\hyper@linkstart##1##2{}%
     \let\hyper@linkend\@empty\citet[#1][#2]{#3}}}
  \newcommandtwoopt{\citeyearads}[3][][]%
    {\href{http://adsabs.harvard.edu/abs/#3}
    {\def\hyper@linkstart##1##2{}%
     \let\hyper@linkend\@empty\citeyear[#1][#2]{#3}}}
\makeatother

\begin{document}

\title{A new massive double-lined spectroscopic binary system: The Wolf-Rayet star WR\,68a}

   \author{A. Collado
          \inst{1,2}
%\fnmsep\thanks{Visiting Astronomer, Complejo Astron\'omico El Leoncito (Argentina) and 
%Cerro Tololo Inter-American Observatory (Chile).},
          R. Gamen\inst{3}
%\fnmsep\thanks{Visiting astronomer, Cerro Tololo
%    Inter-American Observatory (Chile) and CASLEO (Argentina).},
          R. H. Barb\'a\inst{4}
%\fnmsep\thanks{Visiting astronomer, Las Campanas
%            Observatory and Cerro Tololo Inter-American Observatory, Chile.},
\and
          N. Morrell\inst{5}
%\fnmsep\thanks{Visiting astronomer, Las Campanas
%            Observatory, Chile}
              }

   \institute {Instituto de Ciencias Astron\'omicas, de la Tierra y del Espacio (ICATE), CONICET, Avda. Espa\~na 1512 Sur, J5402DSP, San Juan, Argentina\\
\email{acollado@icate-conicet.gob.ar}      
      \and Facultad de Ciencias Exactas, F\'isicas y Naturales-Universidad Nacional de San Juan, San Juan, Argentina
            \and Instituto de Astrof\'isica de La Plata, CONICET,
            Facultad de Ciencias Astron\'omicas y Geof\'isicas, Universidad Nacional de La Plata, Paseo del Bosque s/n, B1900FWA, La Plata, Argentina
                      \and
         Departamento de F\'isica y Astronom\'ia, Universidad de La Serena, Av. Juan Cisternas 1200 Norte, La Serena, Chile
                  \and Las Campanas Observatory, Carnegie Observatories, Casilla 601, La Serena, Chile\\
                  }

   \date{Received; accepted}

\abstract{

Double-lined spectroscopic binary systems, containing a Wolf-Rayet and a
massive O-type star, are key objects for the study of massive star evolution
because these kinds of systems allow the determination of fundamental astrophysical
parameters of their components. 
We have performed spectroscopic observations of the star WR\,68a as part of a dedicated monitoring program of WR stars to discover new binary systems. We identified spectral lines of the two components of the system and disentangled the spectra. We measured the radial velocities in the separated spectra and determined the orbital solution. We discovered that WR\,68a is a double-lined spectroscopic binary with an orbital period of 5.2207 days, very small or null eccentricity, and inclination ranging between 75 and 85 deg. We classified the binary components as WN6 and O5.5-6. The WN star is less massive than the O-type star with  minimum masses of 15$\pm$5 \textit{M}$_\odot$ and 30$\pm$4 \textit{M}$_\odot$, respectively.
The equivalent width of the He\,{\sc ii} $\lambda$4686 emission line shows variations with the orbital 
phase, presenting a minimum when the WN star is in front of the system. 
The light curve constructed from available photometric data presents minima in both conjunctions of the system. 
}
\authorrunning{Collado et al}
\titlerunning{A new massive binary system: The Wolf-Rayet star WR\,68a}
   \maketitle
%
%________________________________________________________________

\section{Introduction}

Observational works dedicated to the study of multiplicity among massive stars,
e.g. \citet{2010RMxAC..38...30B}, %\citet{2011IAUS..272..474S},
\citet{2013A&A...550A.107S}, and many references there in,
provide evidence that the fraction of O+OB systems is high, greater than 50\%.
This situation is not completely replicated by the available data for Wolf-Rayet (WR) stars.
There are about 50 binary systems detected among the galactic WR stars (less than
10\% of the known whole sample), and only 24 of them are known double-lined spectroscopic systems (SB2).
These systems are key objects because they allow the determination of some fundamental
astrophysical parameters for their component stars. Moreover, 
double-lined eclipsing systems allow the determination of
stellar masses and radii in a direct and reliable way, thus providing
strong constraints on stellar physics and evolutionary models.
In this context, the discovery of these kinds of systems is highly relevant.

In an attempt to search for new binary systems among faint WR stars, a spectroscopic monitoring of
southern galactic WR stars is underway. 
This survey was started in 2007 and its main results published to date
are the discovery of the SB2 WR~62a, one of 
the most conspicuous radial-velocity (RV) variable stars in our sample 
\citep[hereafter~Paper~I]{2013A&A...552A..22C}, and WR~35a \citep{2014A&A...562A..13G}.

\object{WR 68a} (SMSNPL13, 
\textit{v}=14.41) was identified as a WR star by \citet{Shara_1999}. 
No other data or analysis on this star were found in the bibliography, thus it was included in our 
survey.

In this paper, we present the first spectroscopic and photometric analysis
of WR\,68a, demonstrating that this is an
SB2 system. % {\bf composed} by a WN- and an O-type component.
The paper is organised as follows: in section 2, we describe the observations 
and data reduction. In section 3, we present the spectral analysis of the 
system's components, the measurement and analysis of radial velocities, and the study of 
the available photometry. In section 4, we summarise our results.

%__________________________________________________________________

\section{Observations and data reduction}

%_________________________________________________ One column table
 
\begin{table*}[t]
\tiny
\caption{Details of the observing material.}
\label{table:1}      % is used to refer this table in the text
\centering                          % used for centering table
\begin{tabular}{c c c c c c c c c c}      % centered columns (10 columns)
\hline\hline\\                 % inserts double horizontal lines 
     Date-Obs.      & \textit{n} &Sp. coverage & Observat.& Telesc.&Spectr.$^{{a}}$& Grating        &Detector& Dispersion &  Resolving $^{{b}}$\\
         UT         &   & [\AA ]     &         &  [m]   &       &[l~mm$^{-1}$]   &        &[\AA ~pix$^{-1}$]& power  \\
\hline
\\

 2007, Mar.  30     & 2 & 3650-6700 & CTIO   & 4    & R-C   & 632 & Loral 3k     & 1.01 & 1300\\
 2007, Apr. 1-3     & 4 & 3650-6700 & CTIO   & 4    & R-C   & 632 & Loral 3k     & 1.01 & 1300\\
 2008, Apr. 19-23   & 7 & 3650-6700 & CTIO   & 4    & R-C   & 632 & Loral 3k     & 1.01 & 1300\\
 2008, Jun. 16      & 1 & 3650-6750 & LCO    & 6.5  & IMACS f/4 & 600 & SITe (Mosaic 1)  & 0.4  &  3300\\  
 2009, Mar. 24-27   & 7 & 4030-5590 & CASLEO & 2.15 & REOSC SD& 600 & Tek1024      & 1.63 & 1000 \\
 2009, Jul. 23      & 1 & 3890-5520 & LCO    & 2.5  & B-C   & 1200& Marconi$\#1$ & 0.79 & 2500\\
 2009, Jul. 26      & 1 & 3540-6700 & LCO    & 2.5  & B-C   & 600 & Marconi$\#1$ & 1.55 & 1200 \\
 2009, Jul. 27      & 1 & 3890-5520 & LCO    & 2.5  & B-C   & 1200& Marconi$\#1$ & 0.79 & 2500\\
 2009, Aug.  14-17  & 3 & 4040-5700 & CASLEO & 2.15 & REOSC SD& 600 & Tek1024      & 1.63 & 1000 \\
 2010, Apr.  10-11  & 2 & 3930-5600 & CASLEO & 2.15 & REOSC SD& 600 & Tek1024      & 1.63 & 1000\\
 2010, Aug.  2-8    & 4 & 3930-5600 & CASLEO & 2.15 & REOSC SD& 600 & Tek1024      & 1.63 & 1000 \\
 2011, Apr.  6-11   & 2 & 3930-5600 & CASLEO & 2.15 & REOSC SD& 600 & Tek1024      & 1.63 & 1000\\
 2013, May 12       & 1 & 3930-5600 & CASLEO & 2.15 & REOSC SD& 600 & Tek1024    & 1.63 & 1000  \\
 \hline   \\                                %inserts single line
\multicolumn{10}{l}{{\bf Notes.} (${{n}}$): Number of spectra taken per run. $^{({a})}$ Details of the spectrographs can be found in the user manuals of the respective observatories.} \\
\multicolumn{10}{l}{$^{({b})}$ The spectral resolving power (R=$\lambda/\Delta\lambda$) were measured using $\Delta\lambda$ as the FWHM of calibration lamp emission lines.}\\
\end{tabular}
\end{table*}

%
%______________________________________________________________

We obtained thirty-six spectroscopic observations of \object{WR 68a} between 2007 and 2013. 
The observations were carried out with the 2.15-m J.~Sahade telescope at the 
Complejo Astron\'omico El Leoncito (CASLEO, Argentina), the 4-m V. Blanco telescope at Cerro Tololo 
Inter-American Observatory (CTIO, Chile), and the 2.5-m du Pont and the 6.5-m Magellan Baade 
telescopes at Las Campanas Observatory (LCO, Chile). 
See Table~\ref{table:1} for a summary of the observations. Immediately after or before the stellar 
integration, at the same telescope position, we obtained compari\-son lamp spectra. 
We reduced the spectra using IRAF\footnote{IRAF is distributed by the National Optical 
Astronomy Observatories, which are operated by the Association of Universities for Research 
in Astronomy, Inc., under cooperative agreement with the National Science Foundation.} 
standard procedures.

\section{Results and discussion}

\subsection{The spectrum of WR 68a}
%\label{disen}

The spectrum of WR\,68a shows noticeable emission lines of helium, carbon, and nitrogen ions, and also  
several absorption lines of hydrogen and helium superimposed onto the emissions.
As we prove below, these absorption lines move in anti-phase with respect to the emissions, 
indicating that WR\,68a is a double-lined WR~+~OB binary system. 

To separate the individual components in the spectrum of WR\,68a, we applied a disentangling method 
similar to that developed by \citet{2006A&A...448..283G}. 
Briefly, the method consists of shifting all the spectra to the common RV for one component and co-adding them so that the features of the other component are diluted. In a second step, the pure spectrum of one component (template) is subtracted to obtain a new RV determination. 
These steps are repeated in an iterative way.
The method was only applied to the CTIO and LCO spectra  because CASLEO data have lower signal-to-noise ratio (S/N), and thus they
introduce noise into the mean spectra of each component of the binary. 

\begin{figure*}[t]
      \sidecaption
   \includegraphics[width=18.5cm]{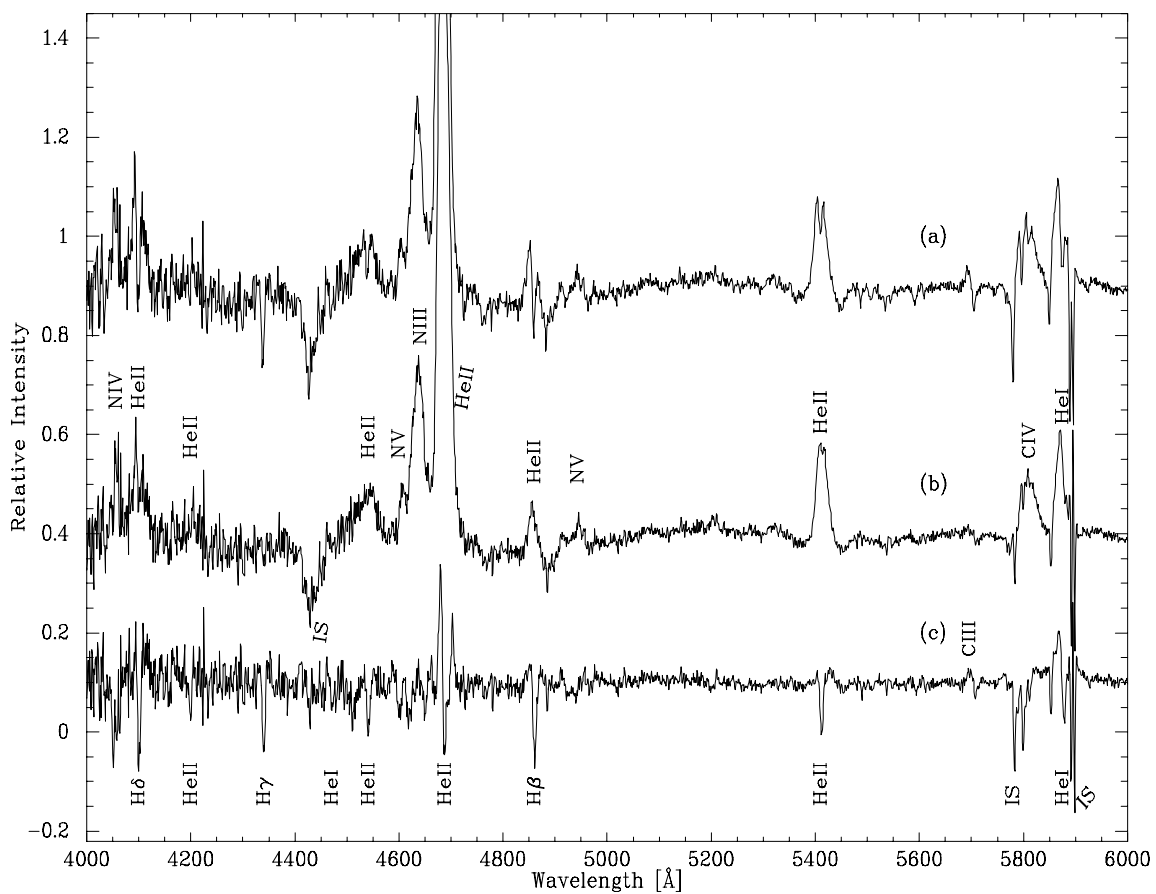}
   \caption{(a) Continuum rectified spectrum of WR\,68a before disentangling; (b) the WR component after remo\-ving the O-type;
(c) the remaining spectrum of the O-type star. The spectra are vertically shifted for comparison.}
\label{disentangling}   
       \end{figure*}

The individual spectra (see Fig.~\ref{disentangling}) were used to perform a more detailed ana\-ly\-sis of both stars. Relative intensities of N\,{\sc iii}, N\,{\sc iv}, He\,{\sc ii}, and 
C\,{\sc iv} emission lines in the spectra indicate a WN6o spectral type \citep[according to the criteria 
given by][]{smi96}\footnote{The ``o'' means that no hydrogen is observed in the spectrum.},
in good agreement with \citet{Shara_1999}.

Absorption lines of H (H$\alpha$, H$\beta$, H$\gamma$, H$\delta$), 
He\,{\sc i} ($\lambda\lambda$4026, 4471, and 5875),  
He\,{\sc ii} ($\lambda\lambda$4542, 4686, and 5411), 
and C\,{\sc iii}~$\lambda$5696~emission were identified in the disentangled spectrum, indicating the 
presence of an O-type star. To determine its spectral type, 
we used the {\sc mgb} code \citep{mgb}, which facilitates a visual comparison with standard O-type 
spectra \citep{Sota_2011,gosss2}.
Thus, we classified the O-type star as O5.5--6.
Because of poor S/N in the blue region of the spectrum and  probable residuals 
of the wings of the H$\delta$ and He\,{\sc ii} $\lambda$4686 emission introduced by the 
disentangling method, we could not 
derive a reliable luminosity class for the secondary spectrum.

\subsection{Measurement and analysis of radial velocities}

The radial velocities (RVs) were determined by means of cross-correlation (using the IRAF {\sc fxcor} 
task) between the separated spectra of each component and the corresponding template. 
For each emission line, an appropriate spectral range was used to determine the 
RV thereof. In the case of the RVs of the absorption lines, we determined a mean RV using the spectral 
ranges of H$\beta$, H$\gamma$, and H$\delta$. 
The individual RVs of the measured emission lines
according to the Heliocentric Julian Dates (HJD) of the observations,
including N\,{\sc iv} $\lambda$4058, the blend N\,{\sc iii} $\lambda\lambda$4634-40-42, 
He\,{\sc ii}~$\lambda\lambda$4686 and 5411, and C\,{\sc iv}~$\lambda$5812 emissions, and the mean RV of 
H$\delta$, H$\gamma$, H$\beta$ absorption lines are listed in Table~\ref{table:2}.

% Table 2 available electrocnically only
\onltab{
\begin{table*}[!ht]
\caption{Observed heliocentric radial velocities of \object{WR 68a}.}             % title of Table
\label{table:2}      % is used to refer this table in the text
\centering                          % used for centering table
\begin{tabular}{c c r r r r r r}        % centered columns (8 columns)
\hline\hline\noalign{\smallskip}       % inserts double horizontal lines

HJD     & Phase & N\,{\sc iv}  & N\,{\sc iii} & He\,{\sc ii} &  He\,{\sc ii} & C\,{\sc iv} & Mean Abs. \\  % table heading 
2450000+&   & 4057.76\AA &  4640.64\AA  & 4685.68\AA  &  5411.52\AA  & 5811.98\AA & H$\delta$-H$\gamma$-H$\beta$\\
        &       & [km~s$^{-1}$] & [km~s$^{-1}$] & [km~s$^{-1}$] & [km~s$^{-1}$] & [km~s$^{-1}$]   & [km~s$^{-1}$] \\

 \hline   \noalign{\smallskip}                     % inserts single horizontal line

4189.83 & 0.99  & -238$\pm$2 & -173$\pm$2 & 58$\pm$1  & 46$\pm$5   & -471$\pm$3 & -113$\pm$3 \\         
4189.85 & 0.99  & -239$\pm$1 & -203$\pm$2 & 41$\pm$2  & 30$\pm$1   & -491$\pm$4 &-158$\pm$3 \\  
4191.86 & 0.38  &54$\pm$2    & -85$\pm$2  & 190$\pm$1 & 274$\pm$2  & -68$\pm$5  &-290$\pm$6 \\ 
4192.89 & 0.58  &            &-450$\pm$1  & -268$\pm$4& -162$\pm$7 & -441$\pm$3 &-110$\pm$6 \\  
4193.67 & 0.72  &-425$\pm$1  & -550$\pm$2 & -309$\pm$3& -301$\pm$3 & -702$\pm$3 &10$\pm$9 \\    
4193.90 & 0.77  &-483$\pm$2  & -542$\pm$1 & -297$\pm$1& -327$\pm$4 & -742$\pm$3 &-15$\pm$9 \\   
4575.90 & 0.94  &-297$\pm$1  & -281$\pm$2 & -64$\pm$1 & -113$\pm$3 & -571$\pm$3 &-36$\pm$3\\    
4576.86 & 0.12  &45$\pm$2    &-17$\pm$1   & 256$\pm$1 & 243$\pm$1  & -213$\pm$3 &-209$\pm$13 \\ 
4577.75 & 0.29  &150$\pm$2   & -38$\pm$2  & 272$\pm$1 & 340$\pm$1  & -32$\pm$5  &-283$\pm$17 \\        
4578.66 & 0.47  &            & -299$\pm$3 & 6$\pm$1   & 103$\pm$1  & -230$\pm$2 &-268$\pm$7 \\  
4578.89 & 0.51  &-188$\pm$4  & -365$\pm$3 & -157$\pm$2& -70$\pm$6  & -249$\pm$2 &-161$\pm$5 \\  
4579.61 & 0.65  &-365$\pm$4  & -545$\pm$2 & -303$\pm$1& -239$\pm$1 & -587$\pm$2 &-32$\pm$10\\   
4579.88 & 0.70  &-478$\pm$5  & -592$\pm$1 & -342$\pm$1& -306$\pm$2 & -693$\pm$3 &-2$\pm$14 \\   
4633.69 & 0.01  &-213$\pm$3  & -160$\pm$3 & 66$\pm$3  & 121$\pm$3  & -411$\pm$7 &-113$\pm$21 \\ 
4914.87 & 0.87  &            & -415$\pm$3 & -177$\pm$3& -270$\pm$2 &            &             \\
4915.85 & 0.06  &            & 18$\pm$1   & 202$\pm$1 & 231$\pm$3  &            &                \\
4915.88 & 0.06  &            & -28$\pm$1  & 214$\pm$1 & 194$\pm$4  &            &                \\
4916.81 & 0.24  &            & -8$\pm$3   & 314$\pm$1 & 308$\pm$1  &            &                \\
4916.84 & 0.24  &            & 33$\pm$4   & 329$\pm$2 & 326$\pm$1  &            &                \\
4917.80 & 0.43  &66$\pm$2    & -114$\pm$3 & 120$\pm$1 & 157$\pm$1  &            &-304$\pm$13     \\      
4917.83 & 0.43  & 49$\pm$2   & -123$\pm$1 & 111$\pm$2 & 148$\pm$1  &            &-283$\pm$12     \\      
5035.65 & 0.00  &-256$\pm$2  & -128$\pm$2 & 93$\pm$1  & 66$\pm$2   &            &-103$\pm$5 \\   
5038.62 & 0.57  &-228$\pm$5  & -392$\pm$1 & -230$\pm$4& -225$\pm$5 & -399$\pm$2 &-76$\pm$9 \\   
5039.64 & 0.77  &-444$\pm$1  & -531$\pm$1 & -283$\pm$1& -338$\pm$3 &            &28$\pm$20 \\    
5059.56 & 0.58  &            & -440$\pm$3 & -194$\pm$3& -131$\pm$3 &            &                        \\
5060.53 & 0.77  &            &            &-304$\pm$2 & -296$\pm$2 &            &                        \\
5060.56 & 0.77  &            &            & -317$\pm$1& -234$\pm$1 &            &                        \\
5296.85 & 0.03  &            &-90$\pm$2   & 135$\pm$2 &            &            &                        \\
5297.81 & 0.22  &            &            & 317$\pm$1 & 367$\pm$1  &            &                        \\
5410.52 & 0.81  &            & -557$\pm$2 & -239$\pm$1&            &            &                        \\
5411.55 & 0.00  &            & -158$\pm$2 & 128$\pm$4 &            &            &                        \\
5413.50 & 0.38  &            &            & 197$\pm$1 & 258$\pm$1  &            &                        \\
5416.51 & 0.95  &            & -156$\pm$2 & -30$\pm$2 &            &            &                \\
5657.88 & 0.19  &            & 73$\pm$2   & 309$\pm$3 & 268$\pm$4  &            &                        \\
5662.74 & 0.12  &            &            & 284$\pm$2 & 216$\pm$1  &            &                        \\
6425.80 & 0.28  &            & 2$\pm$3    &322$\pm$1  & 353$\pm$3  &            &                        \\

\hline                                   %inserts single line
\hline   \noalign{\smallskip}                     % inserts single horizontal line

\textit{FWHM}$^a$ [\AA] & & 14.6 & 30.0 & 22.5 & 21.6   & 34.5          &           \\
\textit{EW}$^a$ [\AA]   & & -4.6 & -8.5 & -31.4&-5.3    & -5.6          &            \\

\hline   \noalign{\smallskip}                     % inserts single horizontal 

\multicolumn{8}{l}{Notes: Radial velocities errors listed in Table are provided by the fxcor task of IRAF. The errors are}\\
\multicolumn{8}{l}{calculated from the asymmetric noise of the cross-correlation function and of the fitted height of the }\\
\multicolumn{8}{l}{peak.}\\
\multicolumn{8}{l}{ $^{(a)}$ The mean FWHM and EW of each emission line were measured in the disentangled spectra.}\\
\end{tabular}
\end{table*}
}% end of onltab

\begin{figure}[t]
\centering
    \includegraphics[width=9cm]{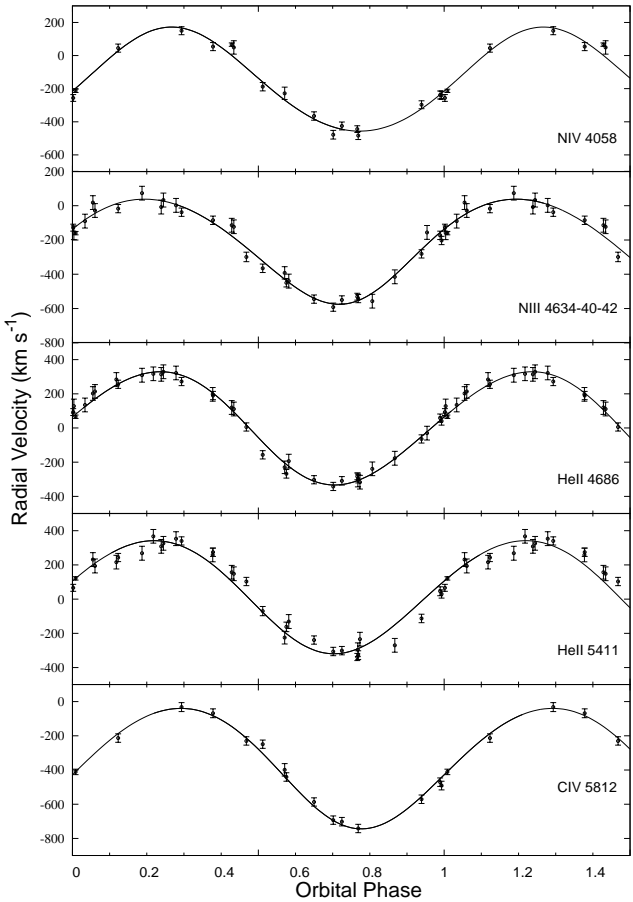}
    \caption{Radial velocities of N\,{\sc iv}, N\,{\sc iii}, He\,{\sc ii}, and C\,{\sc iv} emission lines. Continuous curves represent the orbital solutions in Table~\ref{table:3}.}
\label{vrs}
       \end{figure}

We searched for periodicities in the RVs of the He\,{\sc ii}~$\lambda$4686 emission line, using 
the method developed by \citet{mar80}. This algorithm computes, for each trial period, the variances 
from the best-fitting straight line in each phase interval. 
The most probable period obtained is 5.22 d. 
Then, by means of the {\sc gbart}\footnote{{\sc gbart} is an improved version of the program 
for the determination of the orbital elements for spectroscopic binaries originally 
written by \citet{ber68}, developed by F. Bareilles, and available at 
http://www.iar.unlp.edu.ar/\textasciitilde fede/pub/gbart.} code, we individually fit orbital solutions 
to the RVs for the emission and absorption lines . 
As all lines show a similar periodicity, we adopted the straight mean of the periods obtained 
from each data set, i.e. \textit{P}=5.2207$\pm$0.0005 d, as the binary period.
Thus, with the fixed period, we independently determined the orbital parameters for the emission and the mean of 
the absorption lines,  and we show these in Table~\ref{table:3}.
The orbital solutions for each emission line are illustrated in Fig.~\ref{vrs}.

As expected, each data set presents different systemic velocities, semi-amplitudes, times of periastron 
passage, and maximum RV.
For example, the blend N\,{\sc iii}~$\lambda\lambda$4634-40-42 presents the lowest semi-amplitude and the 
C\,{\sc iv}~$\lambda$5812 emission appears to have the largest. 
This is a common behaviour in WN+OB systems, generally related to the fact that emission lines are
formed in asymmetric (not barycentric) regions at different depths in the expanding envelope 
\citep[see e.g.][]{nie82,nie95}.
    
It is also noteworthy that the RVs of the absorption lines are anti-phased with respect to the emission 
lines, thus indicating that they belong to the O-type component of the system. 
The semi-amplitude of its orbit is lower than that of the WN (regardless of which emission line  is used),
which means that the O-type star is the more massive star in the system.

Eccentricities other than zero are not common in binary systems
containing classical He-burning WN components. Only WR\,97 is known to
have a significant \textit{$e$} (0.1$\pm$0.04, \citet{Gamen_2004}). 
We checked the robustness of the found \textit{$e$} with a test of significance given by \citet{1971AJ.....76..544L}
and it resulted to be spurious in all of the emission lines (they do not reach the 5\% level of significance for accepting eccentric orbits), but are significant in the absorption lines.
This feature should be interpreted with tailored models appropriate to this context of close binaries, 
i.e. heating hemispheres by a hot companion,  asymmetric emission-line forming regions, colliding wind zone, etc.
On the other hand, as is shown below, the light curve presents two dips at 0.5 phase intervals, also pointing to a circular orbit. In the following analysis, a circular orbit is assumed for the WR\,68a binary system.

\begin{table*}[t]
\tiny
\caption{Orbital solutions corresponding to the radial velocities of the N\,{\sc iv}, N\,{\sc iii}, He\,{\sc ii}, and C\,{\sc iv} emission lines, and Mean Abs.}             
\label{table:3}      
\centering     
     
\begin{tabular}{c c c c c c c }     % 8 columns 
\hline\hline       
Parameter       & N\,{\sc iv}&  N\,{\sc iii}&He\,{\sc ii} &He\,{\sc ii} & C\,{\sc iv} &  Mean Abs.\\              
                &4057.76    &   4640.64                 &4685.68     &5411.52     &   5811.98 &H$\delta$-H$\gamma$-H$\beta$\\                 

\hline  \noalign{\smallskip}

\textit{P} [days]       &         \multicolumn{6}{c}{5.2207$\pm$0.0005} \\                                                                                                                  
\noalign{\smallskip}

\textit{V$_0$} [km~s$^{-1}$]          & -159$\pm$7      & -245$\pm$5       & 4$\pm$3       & 22$\pm$5      & -373$\pm$4     & -146$\pm$4\\       
\textit{K} [km~s$^{-1}$]                     &   314$\pm$11     & 307$\pm$7        & 331$\pm$4     & 331$\pm$7     & 352$\pm$7     & 158$\pm$6 \\        
\textit{e}                   & 0.05$\pm$0.03    & 0.08$\pm$0.02    & 0.05$\pm$0.01 & 0.04$\pm$0.02 & 0.06$\pm$0.01 & 0.20$\pm$0.04\\       
\textit{w} [degrees]         &  350$\pm$37      & 206$\pm$17       & 111$\pm$12    & 134$\pm$30    & 157$\pm$19    &       262$\pm$9\\  
\textit{T}$_{\rm Periast}$ [d]$^{*}$& 5036.9$\pm$0.5    &  5039.7$\pm$0.2  & 5038.4$\pm$0.2& 5038.8$\pm$0.4& 5039.4$\pm$0.3& 5038.4$\pm$0.1\\     
\textit{T}$_{\rm RVmax}$ [d]$^{*}$& 5037.1$\pm$0.5      & 5036.7$\pm$0.2   & 5036.8$\pm$0.2& 5036.9$\pm$0.4& 5037.2$\pm$0.3& 5039.4$\pm$0.1 \\   
\textit{T}$_{\rm WR in front}$ [d]$^{*}$& 5038.3$\pm$0.5 & 5038.1$\pm$0.2 &5038.1$\pm$0.2 & 5038.2$\pm$0.4 & 5038.5$\pm$0.3 & 5035.9$\pm$0.1 \\
\textit{a} \textrm{sin i} [\textit{R}$_\odot$] & 32$\pm$1               & 32$\pm$1           & 34$\pm$1      & 34$\pm$1      &  36$\pm$1     & 16$\pm$1\\
\textit{$\sigma$} [km~s$^{-1}$]     & 31.49             & 35.57            & 23.51          & 37.37         & 17.12         & 21\\               
\hline  

{$^{*}$ HJD-2 455 000}\\
%\multicolumn{10}{l}{$^{*}$ +2450000}\\
\end{tabular}
\end{table*}

\begin{figure}[t]
\centering
  \includegraphics[width=9cm]{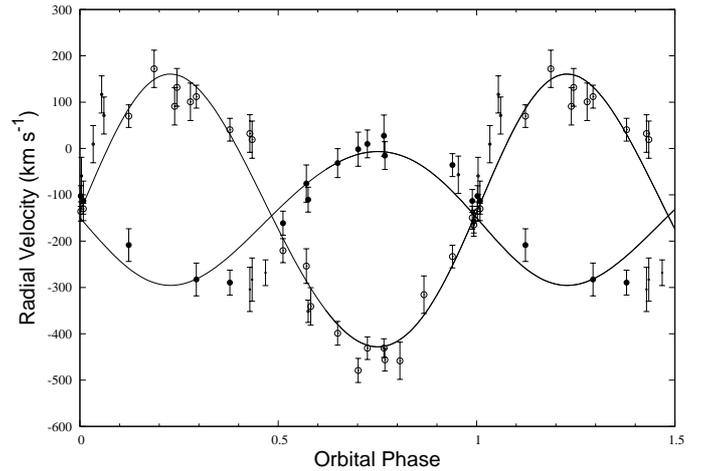}
    \caption{\object{WR 68a} RV orbit represented by the mean of emission lines of nitrogen 
          % N~{\sc iv}~$\lambda$4058 and C~{\sc iv}~$\lambda$5812 emission lines 
for the WN component (open circles) and the mean of H$\beta$, H$\gamma$, and H$\delta$ absorption 
lines for the secondary component (close circles). Points not considered in the solution are plotted with smaller sizes.}
  \label{fig:sb2}
       \end{figure}

To determine the orbital solution of both components simultaneously, 
we assumed that the orbital motion of the WN star is better represented by the mean RVs of the 
nitrogen emission lines, and the motion of the O-type star is better
depicted by the most conspicuous lines in the O-type spectrum.
We shifted the RVs of emission lines  by the difference between the barycentric velocities obtained for the individual solutions to match the lines of the the O-type star component.
We performed an iterative process to determine the orbital solution, rejecting those points whose O-C values were three times larger than the rms of the whole solution in each trial.
The orbital parameters and the RV orbits are shown in Table~\ref{sb2} and Fig.~\ref{fig:sb2}, respectively.

The minimum mass derived for the O-type star is very similar to the theoretical value for an
O~5.5--6~V star, as calibrated by \citet{2005A&A...436.1049M}, 31-33 \textit{M}$_\odot$. 
Therefore, the orbital inclination should be close to 90~deg, and we can expect to observe 
photometric variability (eclipses, for instance).

\begin{table}[t]
\tiny
\caption{Cicular orbital elements of \object{WR 68a} using the mean of emission lines of nitrogen for the WN component and the mean of H$\beta$, H$\gamma$, and H$\delta$ absorption lines for the O-type component.}
           
\label{sb2}      
\centering     
\begin{tabular}{c c}     % 3 columns 
%\hline\hline \noalign{\smallskip} 
    
\hline  \noalign{\smallskip}
\textit{P} [d]                          &         {5.2207 (fixed)}                              \\
\textit{K}$_{\rm WR}$                           &        295$\pm$7    \\
\textit{K}$_{\rm O}$ [km~s$^{-1}$]              &        144$\pm$9    \\
\textit{V}$_0$ [km~s$^{-1}$]                    &        -145$\pm$4    \\
\textit{T}$_{\rm RVmax}$ [d]$^{*}$   &   5039.56$\pm$0.02 \\
\textit{T}$_{\rm WR in front}$ [d]$^{*}$&    5040.86$\pm$0.02 \\
\textit{a}$_{\rm WR}$ \textrm{sin} \textit{i} [\textit{R}$_\odot$]&     33$\pm$1         \\
\textit{a}$_{\rm O}$ \textrm{sin} \textit{i} [\textit{R}$_\odot$] &      15$\pm$1        \\
\textit{M}$_{\rm WR}$\textrm{sin$^{3}$} \textit{i} [\textit{M}$_\odot$]& 15$\pm$5\\
\textit{M}$_{\rm O}$ \textrm{sin$^{3}$} \textit{i} [\textit{M}$_\odot$] & 30$\pm$4\\
\textit{q$_{\rm WR/O}$}                 &         0.49$\pm$0.04     \\
\textit{$\sigma$} [km~s$^{-1}$]         &  32.12 \\
\hline \noalign{\smallskip}  

\multicolumn{2}{l}{\tiny $^*$ HJD-2 455 000}\\
\end{tabular}
\end{table}

\subsection{{\bf Analysis of the emission-line profiles and available photometry}}
\label{ewsection}

\begin{figure}[t]
    \centering
    \includegraphics[width=9.5cm]{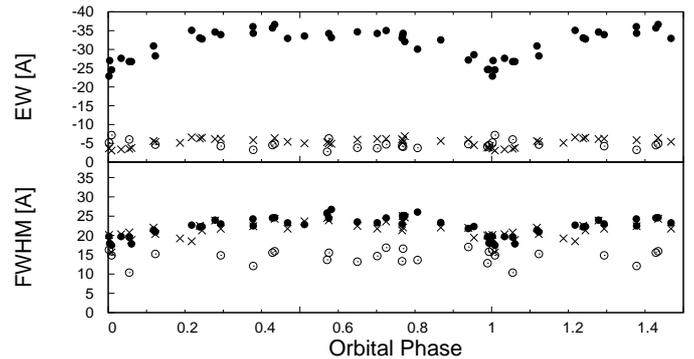}
    \caption{EW and FWHM, as a function of orbital phase, measured in the He\,{\sc ii}~$\lambda$4686 (close circles), 
    N\,{\sc iv}~$\lambda$4058 (open circles), and He\,{\sc ii}~$\lambda$5411 (crosses) emission lines.}   
    \label{ew}
\end{figure}

We analysed the behaviour of the equivalent width (EW) and full width at half maximum (FWHM) of some 
lines at different orbital phases. We found that the EW and FWHM of the He\,{\sc ii}~$\lambda\lambda$4686, 
5411 and N\,{\sc iv}~$\lambda$4058 emission lines present some variability modulated by the orbital period,
which is very noticeable in the He\,{\sc ii} $\lambda$4686 line (see Fig.~\ref{ew}). 
This line becomes fainter around phase $\phi$=0.0, i.e. when the WN star is in the front of the system. 
A similar effect has previously been observed in WR\,62a (see Paper~I)
and in the intriguing LBV HD~5980 \citep{Foellmi_2008}, and it has been explained as due to the emission 
line having an additional non-stellar component, which originates in the colliding wind region.

We also analysed the data collected by the All Sky Automated Survey (ASAS) \citep{pojmanski_2001}. 
We constructed the light curve (see Fig.~\ref{asas}, where we also show the averaged points for 
each bin of $\phi$=0.05), and noted two dips:  one in phase 0.0 (when the WN-type component passes in 
front of the system) and the other, less pronounced, in phase 0.5. 
To prove it is not fortuitous, the ASAS data were searched for periodicities, applying the same algorithm 
used for the RVs. The most likely period found is \textit{$P$}= 5.22 d, which is similar to the spectroscopic 
period (both periodograms are compared in Fig.~\ref{marmuzASAS}).

The similarity between the light curves of WR\,62a and WR\,68a is notorious, 
however, the dip at $\phi$=0.0 in the light curve of WR\,68a is shallower and wider than 
that observed in WR\,62a. 
The resemblance of these light curves to that of V444 Cyg, a
WN5+O6 binary system, is also remarkable (see Figure 2 of \citealt{2011AN....332..616E}). The photometric data of the ASAS 
survey are not suitable for a detailed analysis because of the faintness of WR\,68a. 
Notwithstanding that, we adjusted a simple Wilson-Devinney (WD) model \citep{1971ApJ...166..605W} by means 
of the {\sc phoebe} code \citep{2005ApJ...628..426P}, to the RV of both stars and the photometric data 
together. We adopted the theoretical stellar mass and radii for an O 5.5-6 V star 
from \cite{2005A&A...436.1049M} and adjusted the stellar parameters of the WN6 component. 
We found that the light curve can be reasonably fitted with an orbital inclination ranging between 
75 and 85 deg. These values imply for the WN component a radius between 4.5 and 7.4 \textit{R}$_\odot$, 
comparable with those estimated for the WN5 component of V444 Cyg 
(\cite{2011AN....332..616E} and references therein).

\onlfig{
\begin{figure}
      \centering
     \includegraphics[width=10cm]{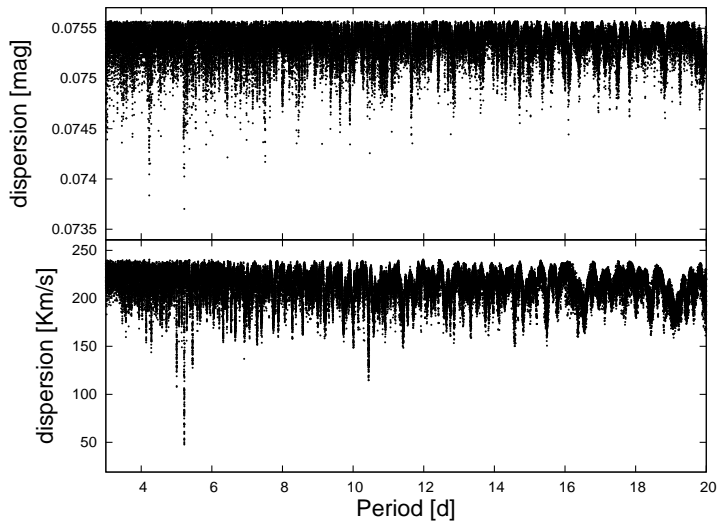}
     \caption{Periodograms obtained for the ASAS data (top) and for the He\,{\sc ii}~$\lambda$4686 
               emission line (bottom).}   
      \label{marmuzASAS}
\end{figure}
}

\begin{figure}[t]
      \centering
     \includegraphics[width=9cm]{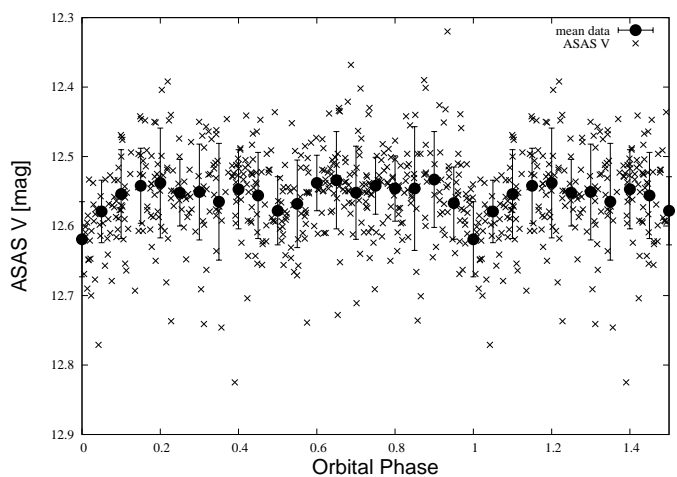}
      \caption{The ASAS V-band photometry of \object{WR 68a}. Crosses
depict the ASAS data, and the circles, the averaged in phase bins of 0.05 with their respective standard deviation.}   
\label{asas}
       \end{figure}

%
%______________________________________________________________

\section{Conclusions}

We have discovered that WR\,68a is a double-lined binary system composed of a WN6 star and an O~5.5-6 
type component in a orbital period of 5.2207 d. 
We found that the O-type star is more massive than the WN component with minimum
masses of 30~$\pm$~4~$\textit{M}_\odot$ and 15~$\pm$~5~$\textit{M}_\odot$ for the O-type and  WN stars, 
respectively. 

We also detected a minimum in the EW of the He\,{\sc ii}~$\lambda$4686 emission when the
WN star is in front of the system, likely due to the eclipse of an additional component of this
line originated in the colliding winds region. 
From the analysis of the available photometric data (ASAS), we found a variability,
i.e. two minima in the light curve during both conjunctions of the system, which we interpreted as eclipses.

WR~68a becomes in an important astronomical target, as accurate multi-band photometric and higher resolution spectroscopic follow-up observations will allow to determine the stellar parameters, as radius and masses, for both massive components. 
WR~68a is thus a target of choice to obtain direct properties of stars in a WR+O system, hence, to obtain valuable constraints to confront evolutionary models of massive stars.

\begin{acknowledgements}
We are grateful to the referee for valuable comments, corrections, and suggestions, which helped improve the paper. 
We thank the directors and staff of CASLEO, LCO, and CTIO for  support and hospitality during our 
observing runs.
CASLEO is operated under
agreement between the Consejo Nacional de Investigaciones Cient\'ificas y
T\'ecnicas de la Rep\'ublica Argentina and the National Universities of La
Plata, C\'ordoba and San Juan.
RHB acknowledges support from FONDECYT Regular Project No. 1140076. 

\end{acknowledgements}

%-------------------------------------------------------------------

\bibliographystyle{aa}
\bibliography{aa_wr62a}

\begin{thebibliography}{22}
\expandafter\ifx\csname natexlab\endcsname\relax\def\natexlab#1{#1}\fi

\bibitem[{{Barb{\'a}} {et~al.}(2010){Barb{\'a}}, {Gamen}, {Arias}, {Morrell},
  {Ma{\'{\i}}z Apell{\'a}niz}, {Alfaro}, {Walborn}, \&
  {Sota}}]{2010RMxAC..38...30B}
{Barb{\'a}}, R.~H., {Gamen}, R., {Arias}, J.~I., {et~al.} 2010, in Revista
  Mexicana de Astronomia y Astrofisica, vol. 27, Vol.~38, Revista Mexicana de
  Astronomia y Astrofisica Conference Series, 30--32

\bibitem[{Bertiau \& Grobben(1968)}]{ber68}
Bertiau, F. \& Grobben, J. 1968, Ric. Astr. Spec. Vat., 8, 1

\bibitem[{{Collado} {et~al.}(2013){Collado}, {Gamen}, \&
  {Barb{\'a}}}]{2013A&A...552A..22C}
{Collado}, A., {Gamen}, R., \& {Barb{\'a}}, R.~H. 2013, \aap, 552, A22

\bibitem[{{Eri{\c s}} \& {Ekmek{\c c}i}(2011)}]{2011AN....332..616E}
{Eri{\c s}}, F.~Z. \& {Ekmek{\c c}i}, F. 2011, Astronomische Nachrichten, 332,
  616

\bibitem[{{Foellmi} {et~al.}(2008){Foellmi}, {Koenigsberger}, {Georgiev},
  {Toledano}, {Marchenko}, {Massey}, {Dall}, {Moffat}, {Morrell}, {Corcoran},
  {Kaufer}, {Naz{\'e}}, {Pittard}, {St-Louis}, {Fullerton}, {Massa}, \&
  {Pollock}}]{Foellmi_2008}
{Foellmi}, C., {Koenigsberger}, G., {Georgiev}, L., {et~al.} 2008, \rmxaa, 44,
  3

\bibitem[{{Gamen}(2004)}]{Gamen_2004}
{Gamen}, R. 2004, Ph.D. Thesis, La Plata University

\bibitem[{{Gamen} {et~al.}(2014){Gamen}, {Collado}, {Barb{\'a}}, {Chen{\'e}},
  \& {St-Louis}}]{2014A&A...562A..13G}
{Gamen}, R., {Collado}, A., {Barb{\'a}}, R., {Chen{\'e}}, A.-N., \& {St-Louis},
  N. 2014, \aap, 562, A13

\bibitem[{{Gonz{\'a}lez} \& {Levato}(2006)}]{2006A&A...448..283G}
{Gonz{\'a}lez}, J.~F. \& {Levato}, H. 2006, \aap, 448, 283

\bibitem[{{Lucy} \& {Sweeney}(1971)}]{1971AJ.....76..544L}
{Lucy}, L.~B. \& {Sweeney}, M.~A. 1971, \aj, 76, 544

\bibitem[{{Ma{\'{\i}}z Apell{\'a}niz} {et~al.}(2012){Ma{\'{\i}}z
  Apell{\'a}niz}, {Pellerin}, {Barb{\'a}}, {Sim{\'o}n-D{\'{\i}}az}, {Alfaro},
  {Morrell}, {Sota}, {Penad{\'e}s Ordaz}, \& {Gallego Calvente}}]{mgb}
{Ma{\'{\i}}z Apell{\'a}niz}, J., {Pellerin}, A., {Barb{\'a}}, R.~H., {et~al.}
  2012, 465, 484

\bibitem[{Marraco \& Muzzio(1980)}]{mar80}
Marraco, H. \& Muzzio, J. 1980, PASP, 92, 700

\bibitem[{{Martins} {et~al.}(2005){Martins}, {Schaerer}, \&
  {Hillier}}]{2005A&A...436.1049M}
{Martins}, F., {Schaerer}, D., \& {Hillier}, D.~J. 2005, \aap, 436, 1049

\bibitem[{{Niemela} {et~al.}(1995){Niemela}, {Cabanne}, \& {Bassino}}]{nie95}
{Niemela}, V.~S., {Cabanne}, M.~L., \& {Bassino}, L.~P. 1995, Revista Mexicana
  de Astronom\'{\i}a y Astrof\'{\i}sica, 31, 45

\bibitem[{Niemela \& Moffat(1982)}]{nie82}
Niemela, V.~S. \& Moffat, A. 1982, Ap.J., 259, 213

\bibitem[{{Pojma{\'n}ski}(2001)}]{pojmanski_2001}
{Pojma{\'n}ski}, G. 2001, in Astronomical Society of the Pacific Conference
  Series, Vol. 246, IAU Colloq. 183: Small Telescope Astronomy on Global
  Scales, ed. {B.~Paczynski, W.-P.~Chen, \& C.~Lemme}, 53--+

\bibitem[{{Pr{\v s}a} \& {Zwitter}(2005)}]{2005ApJ...628..426P}
{Pr{\v s}a}, A. \& {Zwitter}, T. 2005, \apj, 628, 426

\bibitem[{{Sana} {et~al.}(2013){Sana}, {de Koter}, {de Mink}, {Dunstall},
  {Evans}, {H{\'e}nault-Brunet}, {Ma{\'{\i}}z Apell{\'a}niz},
  {Ram{\'{\i}}rez-Agudelo}, {Taylor}, {Walborn}, {Clark}, {Crowther},
  {Herrero}, {Gieles}, {Langer}, {Lennon}, \& {Vink}}]{2013A&A...550A.107S}
{Sana}, H., {de Koter}, A., {de Mink}, S.~E., {et~al.} 2013, \aap, 550, A107

\bibitem[{{Shara} {et~al.}(1999){Shara}, {Moffat}, {Smith}, {Niemela},
  {Potter}, \& {Lamontagne}}]{Shara_1999}
{Shara}, M.~M., {Moffat}, A.~F.~J., {Smith}, L.~F., {et~al.} 1999, \aj, 118,
  390

\bibitem[{Smith {et~al.}(1996)Smith, Shara, \& Moffat}]{smi96}
Smith, L., Shara, M., \& Moffat, A. 1996, MNRAS, 281, 163

\bibitem[{{Sota} {et~al.}(2014){Sota}, {Ma{\'{\i}}z Apell{\'a}niz}, {Morrell},
  {Barb{\'a}}, {Walborn}, {Gamen}, {Arias}, \& {Alfaro}}]{gosss2}
{Sota}, A., {Ma{\'{\i}}z Apell{\'a}niz}, J., {Morrell}, N.~I., {et~al.} 2014,
  \apjs, 211, 10

\bibitem[{{Sota} {et~al.}(2011){Sota}, {Ma{\'{\i}}z Apell{\'a}niz}, {Walborn},
  {Alfaro}, {Barb{\'a}}, {Morrell}, {Gamen}, \& {Arias}}]{Sota_2011}
{Sota}, A., {Ma{\'{\i}}z Apell{\'a}niz}, J., {Walborn}, N.~R., {et~al.} 2011,
  \apjs, 193, 24

\bibitem[{{Wilson} \& {Devinney}(1971)}]{1971ApJ...166..605W}
{Wilson}, R.~E. \& {Devinney}, E.~J. 1971, \apj, 166, 605

\end{thebibliography}

\end{document}